\documentclass[sigconf]{acmart}
\usepackage{stfloats}
\AtBeginDocument{%
  }

\setcopyright{acmlicensed}
\copyrightyear{2024}
\acmYear{2024}
\acmDOI{XXXXXXX.XXXXXXX}

\begin{document}

\title{ESAM: $\underline{\text{E}}$nergy-efficient $\underline{\text{S}}$NN $\underline{\text{A}}$rchitecture using 3nm FinFET $\underline{\text{M}}$ultiport SRAM-based CIM with
Online Learning}








\author{Lucas Huijbregts\textsuperscript{1,2}, Liu Hsiao-Hsuan\textsuperscript{2}, Paul Detterer\textsuperscript{2}, Said Hamdioui\textsuperscript{1}, Amirreza Yousefzadeh\textsuperscript{3,2}, Rajendra Bishnoi\textsuperscript{1}}
\affiliation{
  \institution{\textsuperscript{1}Delft University of Technology, \textsuperscript{2}IMEC, \textsuperscript{3}University of Twente}
  \country{\textsuperscript{1,3}The Netherlands, \textsuperscript{2}Belgium}
  }
\email{{l.c.a.huijbregts,s.hamdioui,r.k.bishnoi}@tudelft.nl, samantha.liu@imec.be, paul.detterer@imec.nl, a.yousefzadeh@utwente.nl}

\renewcommand{\shortauthors}{Huijbregts et al.}

\begin{abstract}
Current Artificial Intelligence (AI) computation systems face challenges, primarily from the memory-wall issue, limiting overall system-level performance, especially for Edge devices with constrained battery budgets, such as smartphones, wearables, and Internet-of-Things sensor systems. In this paper, we propose a new SRAM-based Compute-In-Memory (CIM) accelerator optimized for Spiking Neural Networks (SNNs) Inference. Our proposed architecture employs a multiport SRAM design with multiple decoupled Read ports to enhance the throughput and Transposable Read-Write ports to facilitate online learning. Furthermore, we develop an Arbiter circuit for efficient data-processing and port allocations during the computation. 
Results for a 128$\times$128 array in 3nm FinFET technology demonstrate a 3.1$\times$ improvement in speed and a 2.2$\times$ enhancement in energy efficiency with our proposed multiport SRAM design compared to the traditional single-port design.
At system-level, a throughput of 44 MInf/s at 607 pJ/Inf and 29mW is achieved.
\end{abstract}



\maketitle

\vspace{-3mm}
\begin{acks}
This work is partially funded by the European Union, DAIS (Grant No. 101007273), CONVOLVE (Grant No. 101070374) and \\NEUROKIT2E (Grant No. 101112268).
\end{acks}

\section{Introduction}\label{sec:introduction} 

The demand for Artificial Intelligence (AI) applications to operate on battery-powered Edge devices, such as smartphones, wearables, and diverse Internet-of-Things (IoT) systems, is rapidly growing. These devices handle substantial amounts of data that require processing through AI algorithms. As transmitting raw data can be energy-intensive and introduce higher latency and privacy risks, there is an increasing demand for the (partial) execution of AI applications directly on Edge devices. The primary solution being explored for low-power Edge AI involves neuromorphic computing and the utilization of \textit{Spiking Neural Networks} (SNNs). This approach leverages large-scale parallel operations, integrating advanced techniques like \textit{Computation In-Memory} (CIM), event-based computation, and the reduction of parameter precision \cite{reasonsEdgeAI}. The main challenge for neuromorphic accelerators involves improving the speed and efficiency of \textit{Multiply-and-Accumulate} (MAC) operations. 

In digital CIM, synaptic weights are stored in \textit{Static Random Access Memories} (SRAM), where MAC operations are predominantly conducted through either Adder Trees and their variants \cite{mac-digital-AT1, mac-digital-AT2, mac-digital-AT3, mac-digital-AT4} or \textit{Multiplication In-Memory with sequential accumulation in the SRAM Periphery} (CIM-P) \cite{mac-digital-1row, mac-digital-1row2, mac-digital-1row3, mac-digital-Nrow}. Adder Trees allow enhanced parallelism but come at the price of disrupting the SRAM structure and introducing considerable hardware overhead. In contrast, SRAM-based CIM-P designs minimize hardware overhead and efficiently leverage SNN sparsity, albeit with the trade-off of reduced parallelism in the pre-synaptic neuron dimension. This is because, in typical SRAM digital designs, only one row can be accessed at a time, resulting in the capability for only one pre-synaptic neuron to fire per timestep. Implementations like \cite{mac-digital-Nrow} strive to address this concern through approximate computing, yet this compromises classification accuracy. Another challenge with CIM-P involves spike arbitration; ensuring that only one spike enters the SRAM per timestep. Generally, these arbitration systems are extensive and demand multiple clock cycles per spike \cite{mac-digital-1row}. 
Additionally, on-chip learning is a popular practice for SNNs, enabling adaptability to dynamic environments and training with smaller datasets. On-chip learning is performed much more efficiently with transposable SRAM, facilitating access to cells in both row- and column-wise directions \cite{transposable-barrelShift, mac-digital-1row2, transposable-6T+2T, transposable4, transposable5}. However, the majority of methods either necessitate additional hardware components in the SRAM array, adversely impact cell stability, or lead to slow and high-power Read/Write operations. 
Hence, there is a decisive need for a cost-effective solution for SRAM-based CIM that can deliver a high degree of parallelism without compromising the classification accuracy and which enables efficient on-chip learning.

In this paper, we propose an SRAM-based SNN accelerator using CIM for low-power Edge AI applications, called ESAM. 
To solve the aforementioned problems with the state-of-the-art we implement:
\begin{itemize}
\vspace{-1mm}
    \item Enhanced pre-synaptic neuron parallelism through novel multiport SRAM cells for synaptic weight storage, each equip-ped with multiple decoupled Read ports;
    \item A novel multiport Arbiter circuit designed to assign input spikes to their designated port channels;
    \item Transposable Read/Write access to the multiport SRAM array, facilitating online learning.
\vspace{-1mm}
\end{itemize}

The multiport SRAM cell as well as Arbiter circuit and other peripheral circuits are designed using IMEC's 3nm FinFET technology. Simulation results for a 128$\times$128 array show that our proposed SNN architecture can improve SNN computation speed and energy-efficiency by $3.1\times$ and $2.2\times$ respectively, in comparison to the standard single port SRAM design. Employing the proposed design at system-level and executing a handwritten digit classification application achieves a throughput of 44 MInf/s at just 607 pJ/Inf, while consuming 29 mW of power.

The rest of this paper is organized as follows: Section \ref{sec:background} presents the basics of SNNs, as well as the concept of Transposable crossbar memories. Section \ref{sec:proposed} presents the proposed accelerator architecture with different SRAM cell options, after which Section \ref{sec:results} does evaluations. Finally, Section \ref{sec:conclusion} concludes the paper.
\section{Background}\label{sec:background}

\begin{figure*}[t]
  \centering
  \begin{tabular}{ c c @{\hspace{20pt}} c}
    \includegraphics[width=.8\columnwidth]{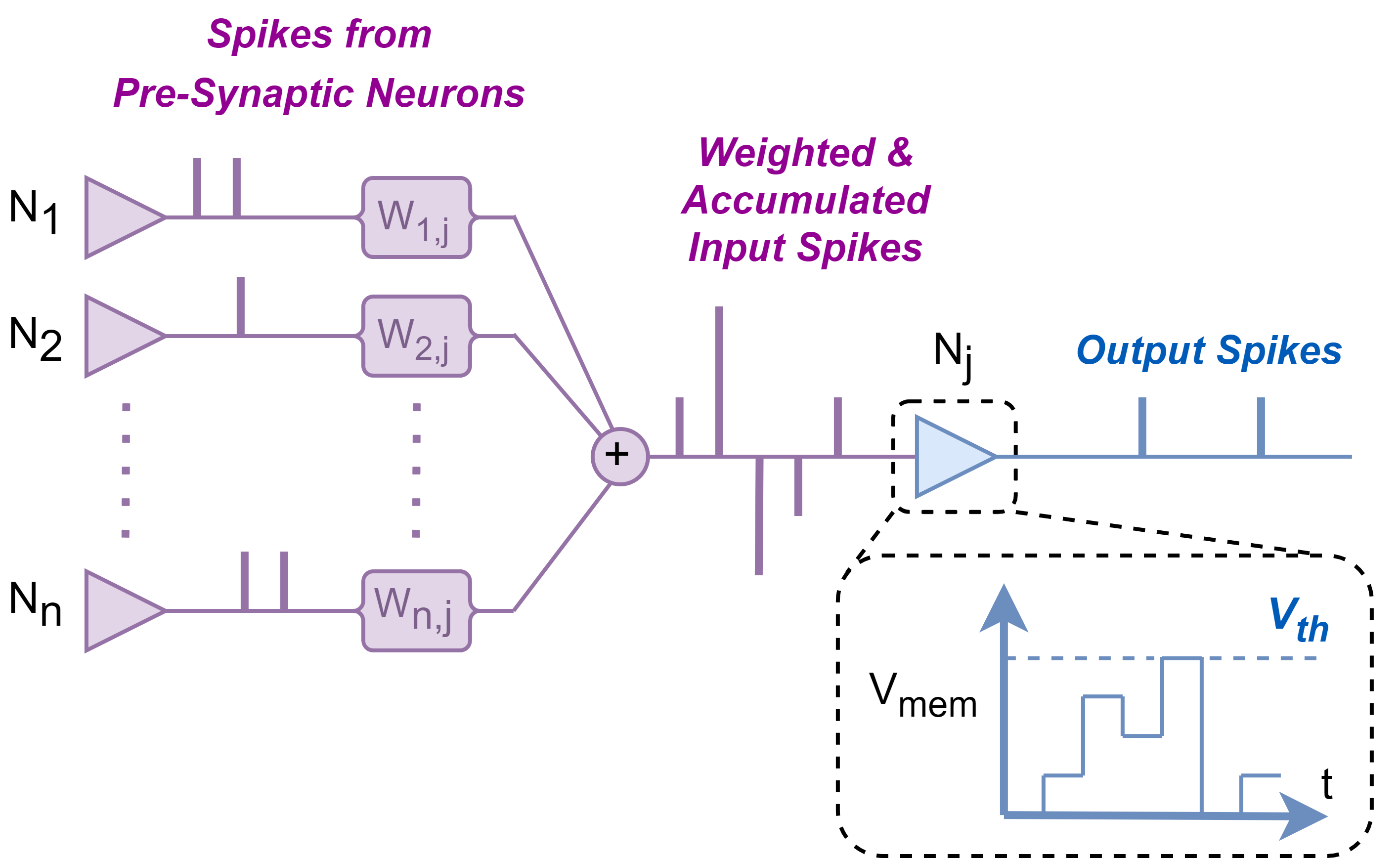} &
    \includegraphics[width=.5\columnwidth]{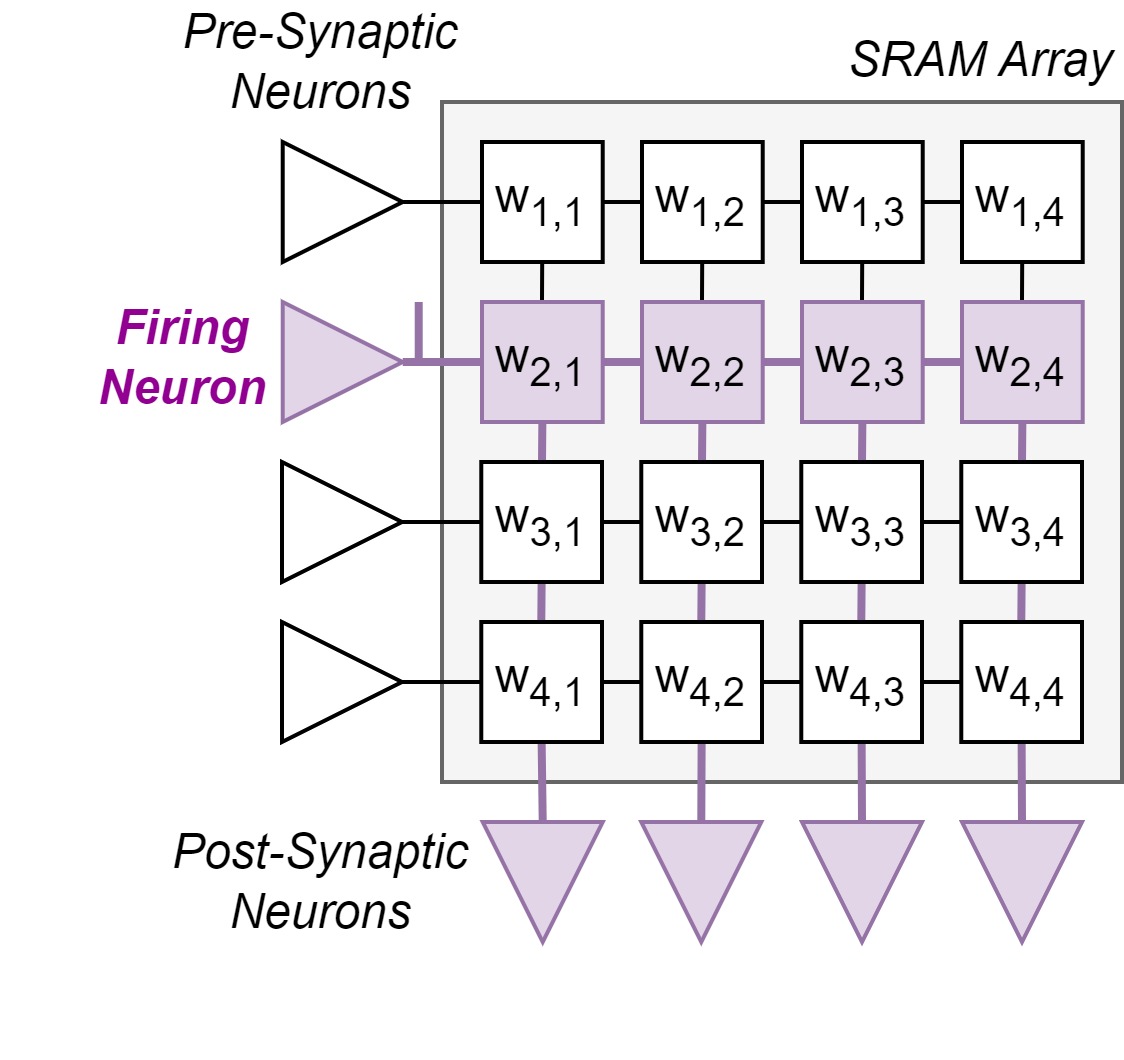} & 
    \includegraphics[width=.5\columnwidth]{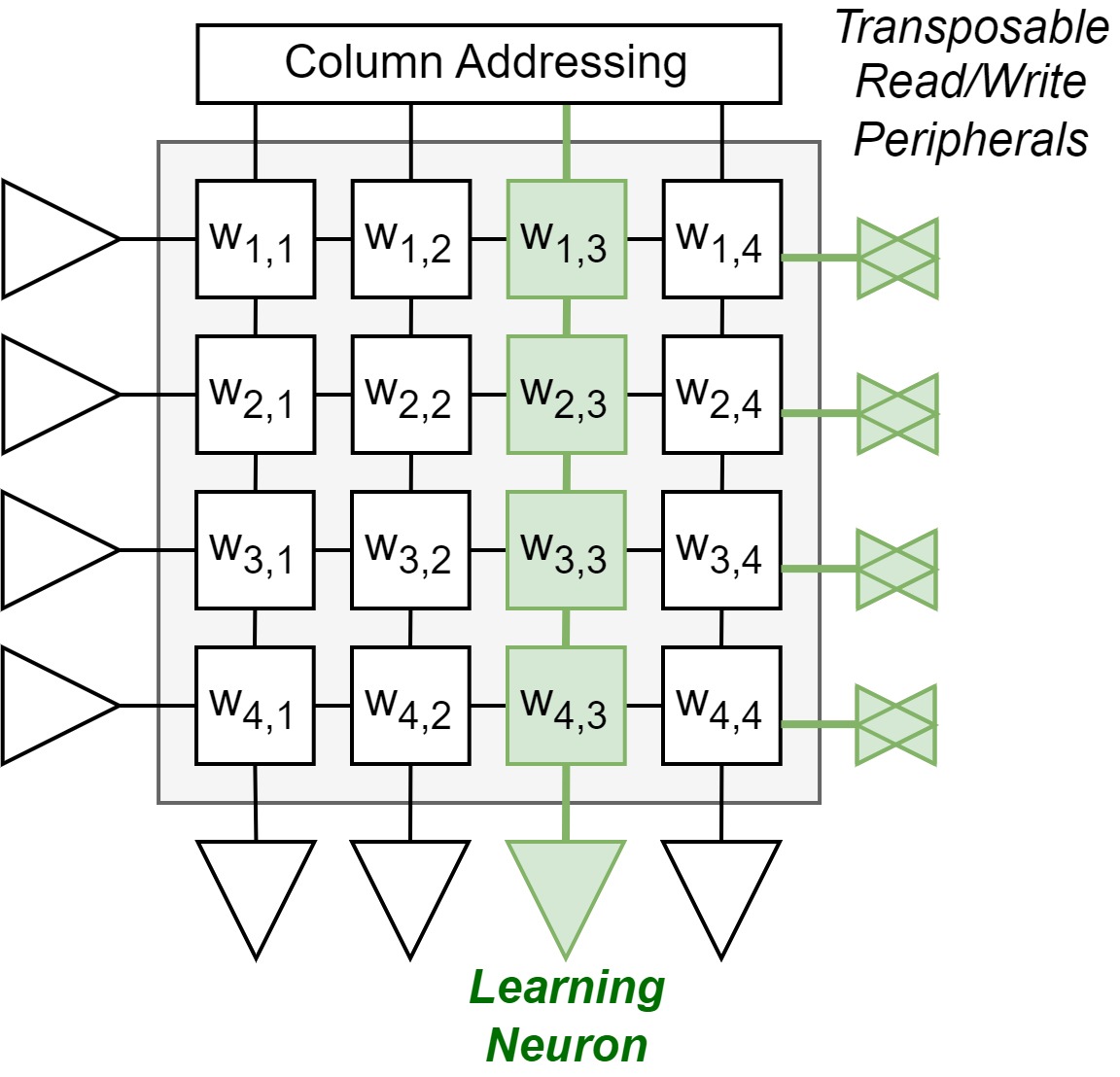}\\
    \small (a)  Illustration of SNN working principle &
      \small (b) Row-wise access for Inference &
      \small (c) Column-wise access for learning
  \end{tabular}


  \caption{Illustration of working principle of Spiking Neural Network, mapping to memory crossbar and traditional row-wise SRAM access versus Transposable column-wise SRAM access.}
  \label{fig:B-SNN-mapping}
\end{figure*}

\subsection{Digital In-memory for Spiking Neural Networks }
\textit{Spiking Neural Networks} (SNNs) represent a significant advancement from traditional Neural Networks \cite{SNN-3rd-gen} by introducing a more biologically inspired model, where neurons communicate through discrete spikes, enhancing the temporal aspects of information processing and enabling more efficient modeling of complex patterns in dynamic data. These spikes contribute to an increased information-carrying capacity and facilitate event-based computing, thereby minimizing energy consumption. 
Figure \ref{fig:B-SNN-mapping}(a) shows a visual example of an SNN. The Figure shows an \textit{Integrate-and-Fire} (IF) neuron, which accumulates weighted spikes as its membrane potential ($V_{mem}$) and sends out a spike when $V_{mem}$ exceeds its threshold potential; $V_{mem} \geq V_{th}$. The digital implementation of such SNNs offers advantages including enhanced energy efficiency, parallel processing capabilities, and compatibility for hardware acceleration, rendering them particularly suitable for real-time, low-power applications in domains like Edge computing. In digital \textit{Computational-In-Memory} (CIM), synaptic weights are mapped to a memory crossbar, as shown in Figure \ref{fig:B-SNN-mapping}(b). Here, \textit{Multiply-and-Accumulate} (MAC) operations are mainly conducted through Adder Tree-based designs \cite{mac-digital-AT1, mac-digital-AT2, mac-digital-AT3, mac-digital-AT4} or through \textit{sequential accumulation in the SRAM Periphery} (CIM-P) \cite{mac-digital-1row, mac-digital-1row2, mac-digital-1row3, mac-digital-Nrow}. 

\subsection{Transposable SRAM for On-Chip Learning}
On-chip learning is a prevalent practice in SNNs, allowing the network to continuously learn even after deployment, ensuring adaptability to changing environments. 
For efficient on-chip learning, it is crucial to have access to the synapse weights in both the pre-synaptic and post-synaptic dimensions \cite{onchiplearning, rostami2022prop, binarySTDP}. Inference necessitates reading in the pre-synaptic dimension, corresponding to a memory row, as indicated in Figure \ref{fig:B-SNN-mapping}(b). Contrarily, learning in SNNs typically occurs when particular conditions arise in the post-synaptic neuron. Hence, weight updates should be applied to all synapses preceding the learning neuron, associated with a memory column, as depicted in Figure \ref{fig:B-SNN-mapping}(c). Overall, row-wise Read access and column-wise Read/Write access are essential for SNN Inference and on-chip learning, respectively. SRAM with access in both the row-wise and column-wise directions is called \textit{Transposable} SRAM.
\section{Proposed ESAM Technique for CIM architecture}\label{sec:proposed}

In this section, we present our proposed SNN accelerator. We first provide an overview of the complete ESAM (Energy-efficient SNN Architecture for Multiport SRAM) system. Subsequently, we provide a more detailed explanation of the innovative subsystem components: the bitcell design, the Arbiter, and the Neuron design.

\subsection{Architecture Overview} \label{sec:proposed:overview} 

As discussed, SRAM-based Computational-In-Memory (CIM) employing Adder Trees to perform the MAC operation results in substantial hardware overhead. Conversely, CIM-P designs reduce hardware overhead and efficiently leverage the sparsity of SNNs; however, they suffer from significantly lower parallelism. The issue lies in the fact that conventional SRAM-based CIM-P designs access memory rows one by one to execute the MAC operation. To enhance the parallelism in the CIM-P architecture, we enable multiple simultaneous accesses to the array through novel multiport processing configurations. This improvement requires addressing input encoding, considering the random nature of input spikes, and managing the accumulation process during MAC operations. To tackle these aspects, we introduce an innovative Arbiter-based encoding scheme and a novel Neuron design capable of handling multiple accesses. Additionally, these multiport capabilities simplify the Transposable arrangements necessary for on-chip learning.

\begin{figure}[t]
    \centering
    \includegraphics[width=\columnwidth]{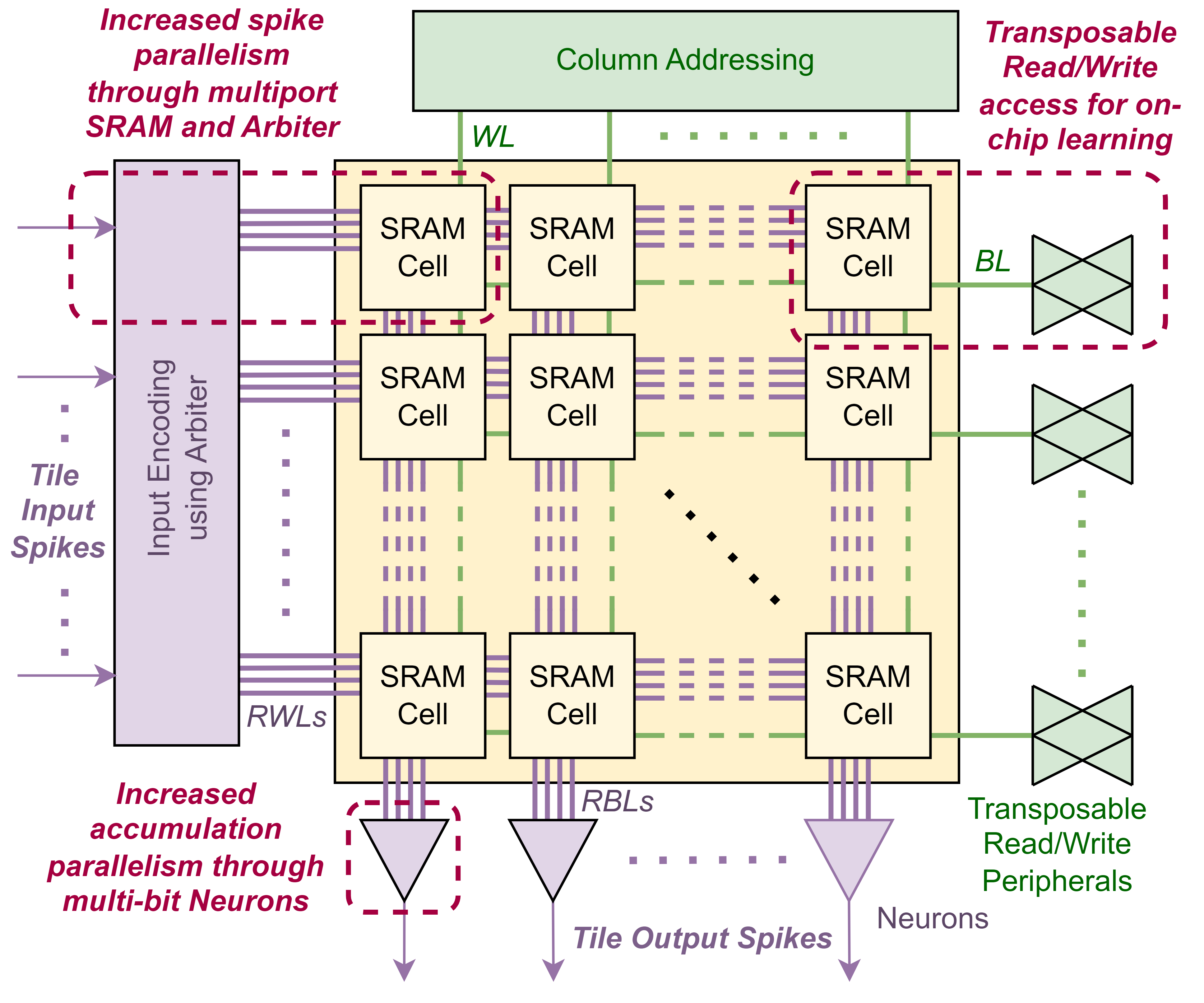}
    \vspace{-7mm}
    \caption{Overview of proposed Macro Architecture. Green indicates the Transposed Read/Write access, with the Inference Read access in Purple.}
    \label{fig:archOverview}
    \vspace{-5mm}
\end{figure}

Figure \ref{fig:archOverview} presents a Macro-level overview of the architecture, illustrating the composition of a single CIM-P Tile. Each Tile executes the MAC operation, involving incoming input spikes and the synapse weights stored in the SRAM array. Multiple Tiles can be integrated to constitute a multi-layer neural network. To create a fully connected neural network, Tiles can be cascaded directly, and spikes are transmitted fully in parallel as binary pulses between the Tiles, negating the need for decoding or routing the spikes. Each Tile contains three novel components, highlighted in Figure \ref{fig:archOverview}: the Arbiter, the SRAM cells, and the Neuron Array. 
The Arbiter processes requests for sending spikes and accommodates up to \textit{p} ($p=4$ in Figure \ref{fig:archOverview}) spike requests per clock cycle, resulting in the activation of corresponding Inference Wordlines (RWLs, depicted in purple). The Read outputs are then conveyed on the Inference Bitlines (RBLs), which then proceed to the Neurons. At the Neurons, these outputs are added together in parallel, after which they determine changes in the membrane potentials $V_{mem}$. Moreover, as depicted in Figure \ref{fig:archOverview} in green, the Transposed Wordlines (WL) and Bitlines (BL) enable column-wise Read/Write access to the SRAM array. This notably decreases the number of operations required to update a column of synapse weights, thereby enhancing the overall efficiency of on-chip learning.

\subsection{Transposable Multiport SRAM cell-based Synapse }

To enhance parallelism and improve the throughput of the architecture, we implement increased accessibility through a multiport SRAM cell design featuring several Read ports capable of processing multiple spikes in parallel. Furthermore, to enhance the efficiency of online learning, we require additional column-wise accessibility. Therefore, to fulfill these requirements, we propose a new multiport cell design that includes one dedicated column-wise Read/Write port and several row-wise Read ports.

\begin{figure}[t!]
  \centering
  \begin{tabular}{ c @{\hspace{20pt}} c }
    \includegraphics[width=.45\columnwidth]{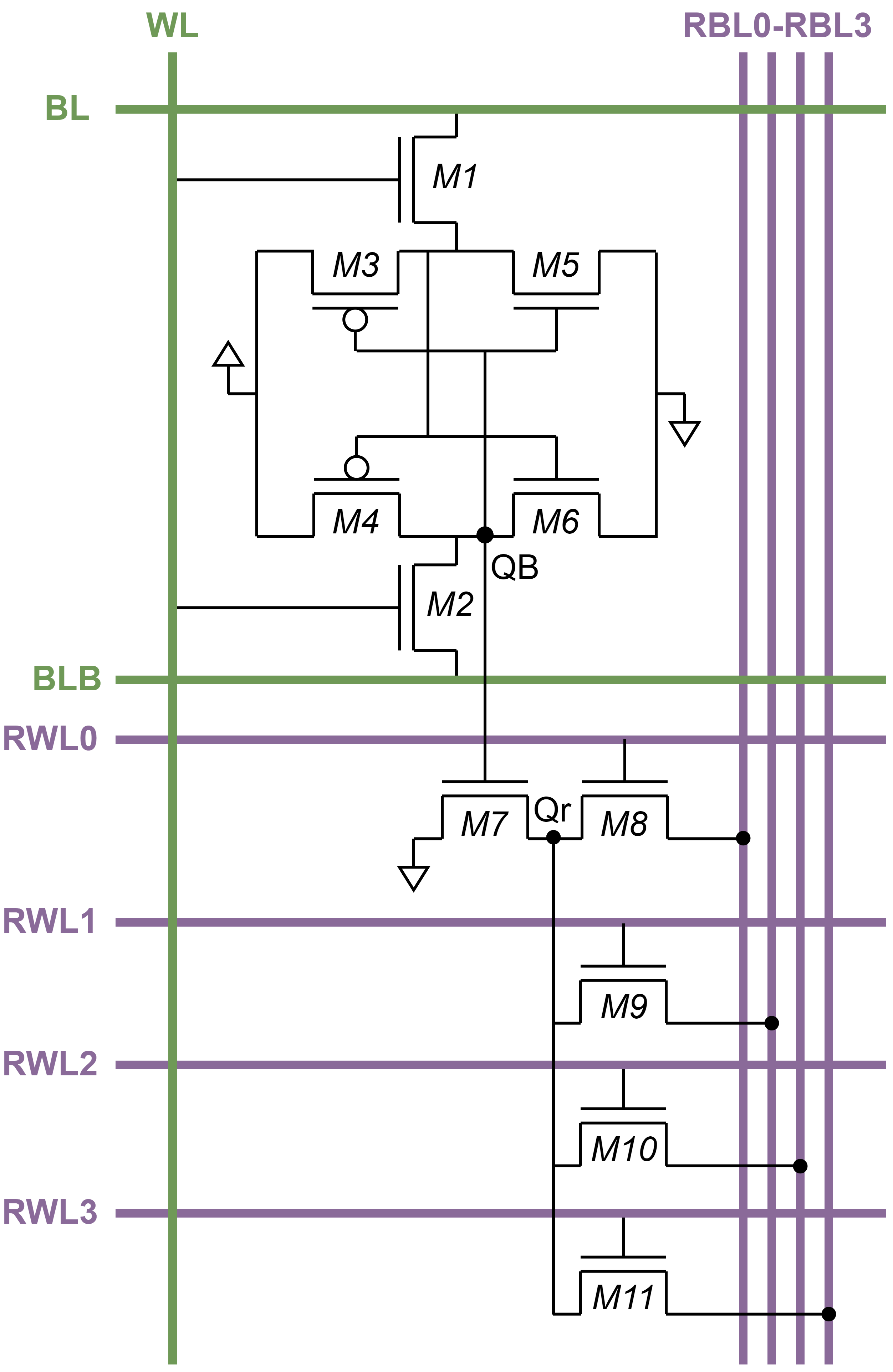} &
      \includegraphics[width=.35\columnwidth]{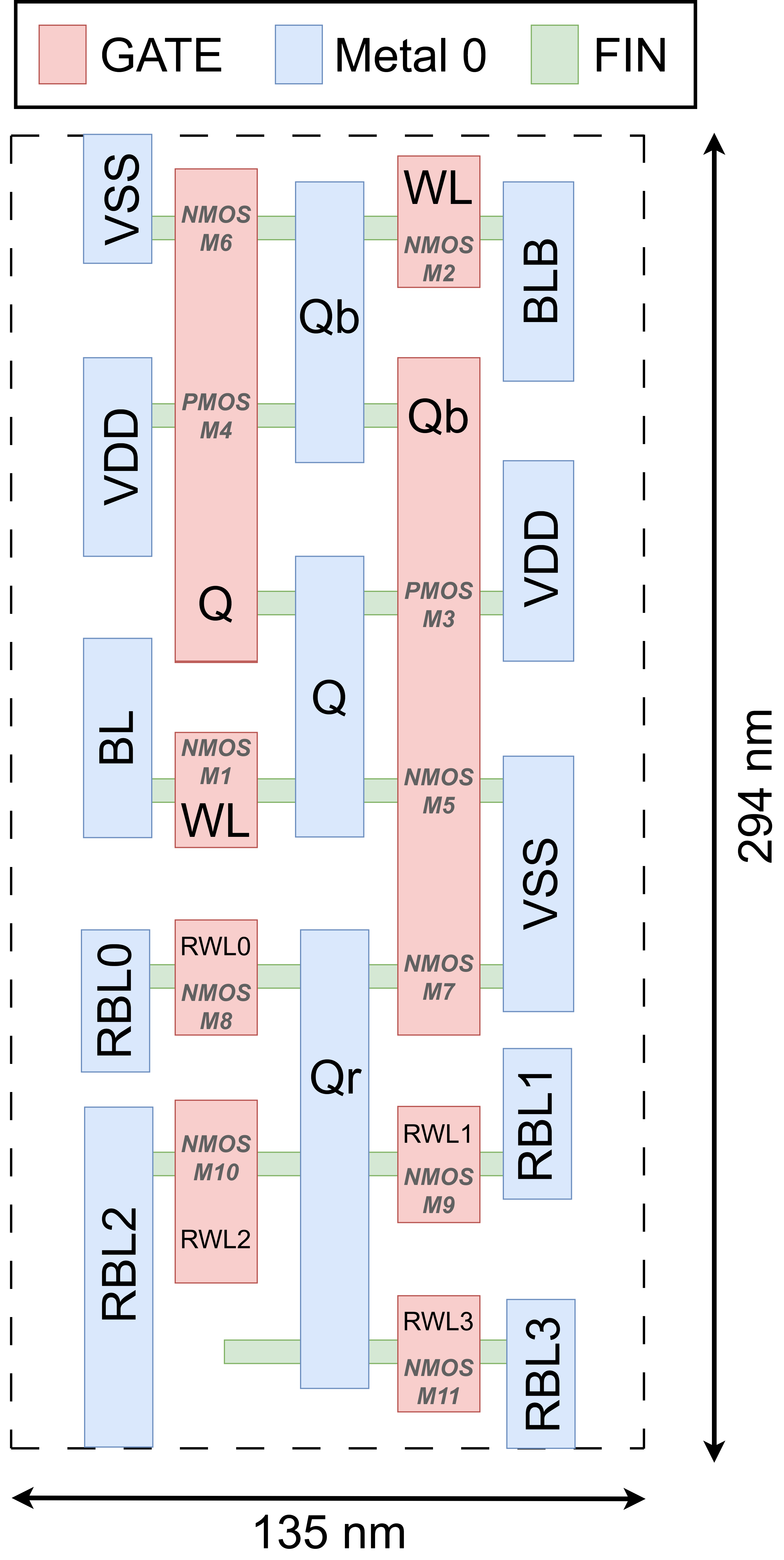} \\
    \small (a) Cell schematic &
      \small (b) Cell layout (using IMEC's \\ & 3nm FinFET)
  \end{tabular}


  \caption{Schematic and layout of proposed five-Read and one-Write (1RW+4R) SRAM Cell.}
  \label{fig:proposedCell}
  \vspace{-5mm}
\end{figure}

Figure \ref{fig:proposedCell}(a) shows the schematic of our proposed one-Read/Write and four-Read (1RW+4R) SRAM cell. The original 6T SRAM cell, comprising transistors M1-M6, remains unchanged; however, in contrast to common convention, its Word Line (WL) runs vertically, while its Bit Line (BL) and Bit Line Bar (BLB) run horizontally, enabling column-wise Read/Write access. M7 provides access to the cell content via the inverted cell content node QB. Through M7, Qr is discharged to VSS if QB is `1', meaning it effectively mirrors the content Q of the cell. M8-M11 then provide access to Qr to one of the added Bitlines RBL0-RBL3, depending on which of the added Wordlines RWL0-RWL3 is set to `1'. To perform a Read operation, an added Bitline, RBLx, is precharged to $V_{prech}$. When the corresponding WLx is driven to `1', the corresponding access transistor conducts. Depending on QB, RBLx is then either kept at $V_{prech}$ or discharged to VSS, which is then sensed.

By connecting M7 only with its Gate to the content of the cell, a decoupled port is formed. This minimizes the influence of the added ports to the stability of the cell. It also enables us to scale $V_{prech}$ to lower values than VDD with almost negligible impact on the cell stability. This results in power savings at the cost of slower precharging. 
 Figure \ref{fig:proposedCell} presents the 1RW+4R cell, which has four dedicated decoupled Read ports and one standard Read/Write port. Similar to this, we have also designed a 1RW+1R, 1RW+2R, and 1RW+3R, each having one, two, and three decoupled Read ports, respectively, alongside one standard Read/Write port. These cells can be inferred from the structure in Figure \ref{fig:proposedCell}(a) by removing Bitlines, Wordlines, and access transistors. The layout of the 1RW+4R cell is shown in Figure \ref{fig:proposedCell}(b). 

To sense the BL/BLB of the Transposed port, we have employed the traditional voltage-based differential Sense Amplifier and used row-muxing by a factor of four to match the SRAM row pitch. For RBL0-RBL3 single ended Reading, we have employed cascaded inverter-based Sense Amplifiers to be able to match the pitch of the SRAM columns, which deliver a slightly slower readout result than traditional Sense Amplifiers. 

\subsection{Arbiter Design} 

\begin{figure}[t!]
    \centering
    \includegraphics[width=0.8\columnwidth]{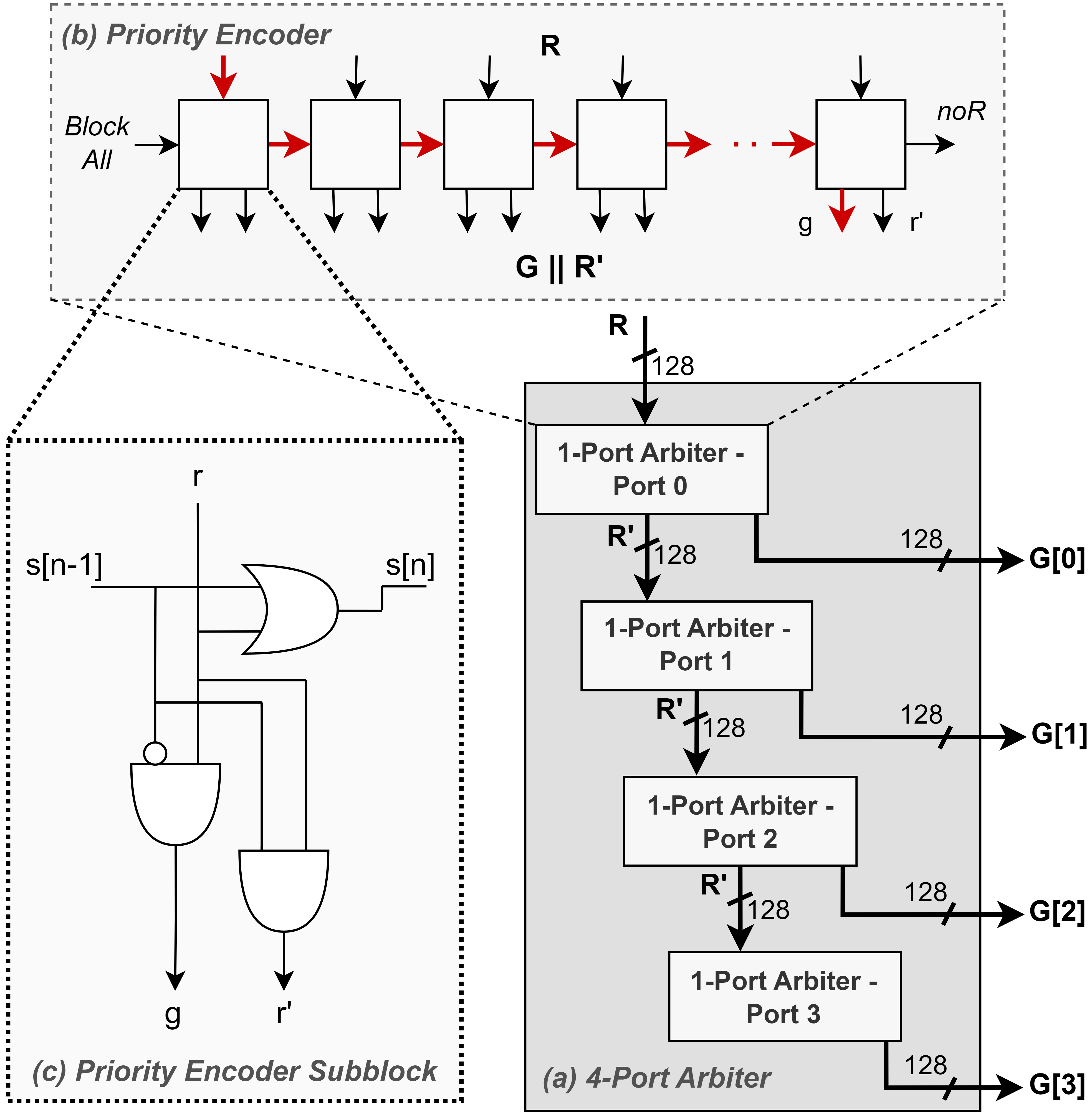}
    \caption{The proposed logic-based 4-Port Arbiter based on four cascaded 1-Port Arbiters.}
    \label{fig:prioEncoder}
    \vspace{-4mm}
\end{figure}

The Arbiter design is implemented to manage spike requests to maximize port utilization and process a higher number of spikes to enhance the overall parallelism of the CIM-P architecture. 
The Arbiter takes a spike Request vector $R$ as input. The vector contains `1's which indicate for which SRAM Wordlines there are pending spike requests. Figure \ref{fig:prioEncoder}(a) shows the structure of the 4-port Arbiter, built out of four cascaded 1-port Arbiters. The 1-port Arbiter is implemented as a Fixed Priority Encoder for this work. This Priority Encoder is highlighted in Figure \ref{fig:prioEncoder}(b). It is built of a string of identical logic-based subblocks, highlighted in \ref{fig:prioEncoder}(c).

The Priority Encoder takes $R$ and selects the leftmost `1' in the vector, creating one-hot Grant vector $G$. The signal $s[n]$ is used to block any requests further to the right of the selection. If $R$ does not contain any spike, $noR$ is made `1'. The vector $R'$ is the same as $R$ except the selected spike is masked out. These non-granted spikes $R'$ are passed to subsequent 1-port Arbiters in a cascaded fashion to extend the system to multiple ports, generating multiple $G$-vectors within a single clock cycle.

In principle the Priority Encoder can be used as an Arbiter by itself, as shown in Figure \ref{fig:prioEncoder}. However, for a full SRAM array of $>128$ rows, this results in an excessively long critical path inside the Priority Encoder, highlighted in Red in Figure \ref{fig:prioEncoder}(b). To mitigate this, in practice the 1-port Arbiter is not implemented as a singular Priority Encoder, but instead by combining multiple shorter Priority Encoders in a tree structure. Multiple shorter base Priority Encoders process the actual request vector $R$, while higher-level Priority Encoders in turn arbitrate amongst these base Priority Encoders. Note that both the base and higher-level Priority Encoders still utilize the same structure as presented in Figure \ref{fig:prioEncoder}(b). At the cost of $8.0\%$ area overhead this reduces the critical path from \textgreater$1100 ps$ to \textless$800 ps$ for the 128-wide, 4-port Arbiter.

\subsection{Neuron Design}

\begin{figure}[t!]
    \centering
    \includegraphics[width=0.9\columnwidth]{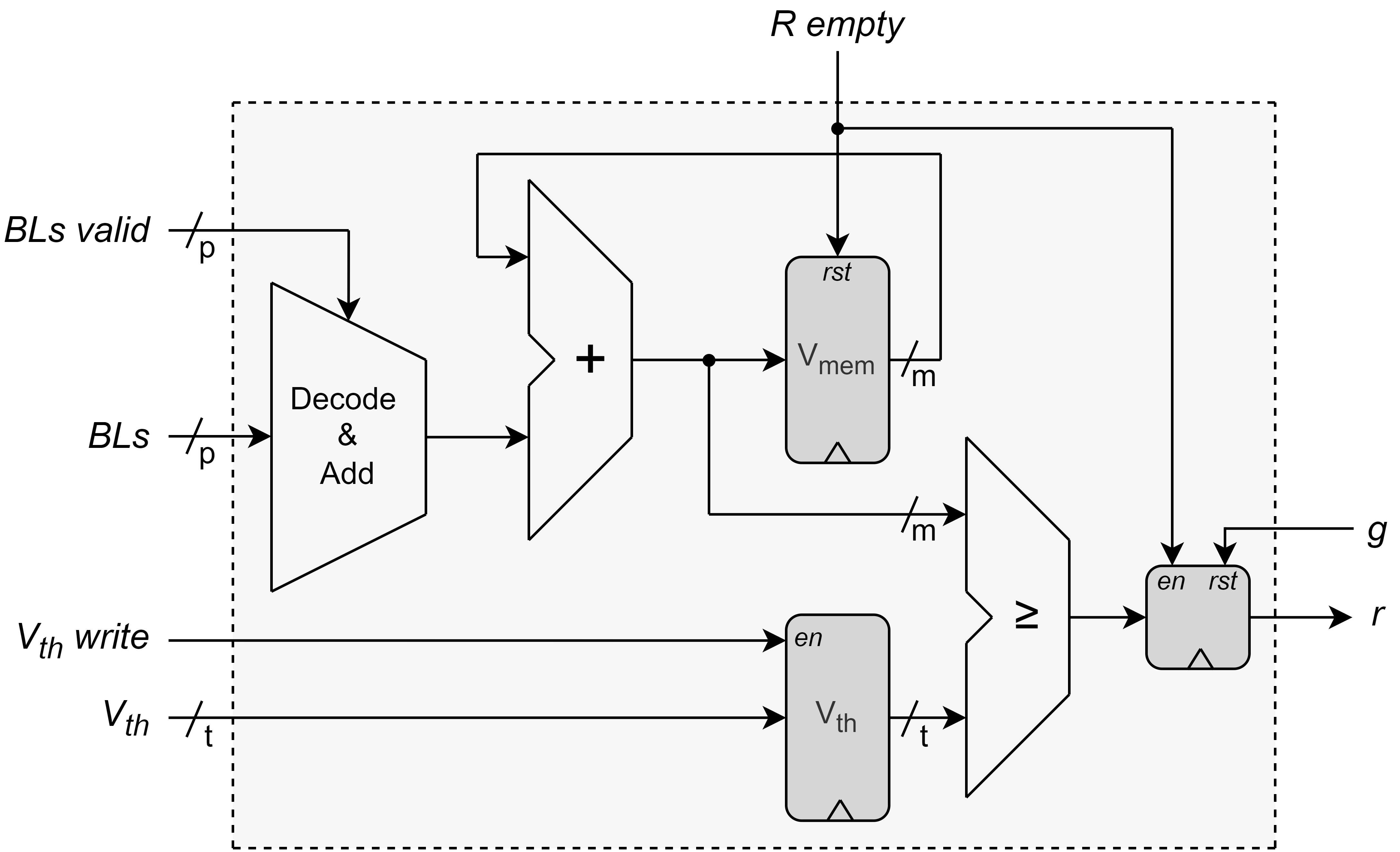}
    \caption{Proposed Neuron for ESAM Architecture.}
    \label{fig:neuronDiagram}
    \vspace{-5mm}
\end{figure}

Figure \ref{fig:neuronDiagram} shows a high-level overview of the Neuron subcomponent. For this work an Integrate-and-Fire (IF) neuron was chosen, as the test setup involves a time-static classification task. The Neurons take in the sensed data from the $p$ multiport Bitlines of the SRAM arrays, represented by \{`1'/`0'\}. For every Bitline, a validity flag is used to indicate which memory ports were actually used in every clock cycle. This ensures an unused port is not erroneously read as a `1' and added to the membrane potential. The valid Bitlines are decoded to \{`+1'/`-1'\}, added together, and then added to the $m$-bit $V_{mem}$ register. The $V_{th}$ of every individual Neuron is stored in its own $t$-bit register. When all input spikes of a Tile have been served by the Arbiter, \textit{R\_empty} becomes `1', which enables the comparison of $V_{mem}$ and $V_{th}$. If $V_{mem} \geq V_{th}$, Neuron output register $r$ is set to `1' to indicate the request to send a spike to the subsequent Tile. Simultaneously, $V_{mem}$ is reset to zero to start accumulating spikes again. If the Neuron's spike request $r$ is granted ($g=$ `1'), $r$ is reset to `0'.

\section{Results}\label{sec:results}
\subsection{Experimental Setup}

Table \ref{tab:setup} summarizes the experimental setup utilized in this work. Note that in order to improve the SRAM Write operation, the \textit{Negative BL} (NBL) assist technique \cite{SamanthaSRAM-TED2022} is used, which creates a lowered voltage ($V_{WD} < VSS$) on the complementary Bitline to force the cell to the desired state. This technique is necessary due to the high parasitics at smaller technology nodes. The required $V_{WD}$ is determined for various array sizes, and if it is necessary for $V_{WD}$ to be $<-400 mV$, the array size is considered non-valid for implementation due to low expected yield \cite{SamanthaSRAM-TED2022}. This restriction limits the array size to $\leq 128$ rows and columns for all cell designs.

The synthesis results, combined with the SRAM Macro outcomes, are utilized to simulate the network on a spike-by-spike basis in Python and determine the timing, power, and energy at the system-level.

\begin{table}[b!]
\centering
\caption{Experimental Setup Details}
\vspace{-4mm}
\label{tab:setup}
\resizebox{1.0\columnwidth}{!}{
\begin{tabular}{l l}
\hline
\textbf{Parameter} & \textbf{Specification}        \\ \hline\hline
Technology Node    & IMEC 3nm FinFet                    \\ \hline
Supply Voltage     & $700 mV$ ($V_{prech} = 500mV$, for single-ended)                        \\ \hline
Simulation         & Cadence Spectre               \\ \hline
Synthesis          & Cadence Genus                 \\ \hline
Parasitics Extraction & \begin{tabular}[c]{@{}l@{}}Calibre PEX + Line Geometries and\\ Node Datasheets\end{tabular} \\ \hline
Process Variation     & $\pm 3 \sigma$                \\ \hline
SRAM Target Cell   & Worst-case Cell/Row/Column    \\ \hline
\end{tabular}}
\end{table}

\subsection{Circuit-Level Evaluation} 

From layout, we have observed that the area of standard 6T is $0.01512 \mu m^2$ \cite{SamanthaSRAM-SPIE-Area}. 
The areas of the 1RW+1R, 1RW+2R, 1RW+3R and 1RW+4R cells are $1.5\times$, $1.875\times$, $2.25\times$ and $2.625\times$ larger respectively. We explored the possibility of adding 5+ ports, but only 4 Bitlines could match the pitch of the 4-port cell. Adding another port would require further widening of the cell, increasing the area by 87.5\% of the 6T cell, making it too area-inefficient. 

\begin{figure}[b!]
    \vspace{-4mm}
    \centering
    \includegraphics[width=\columnwidth]{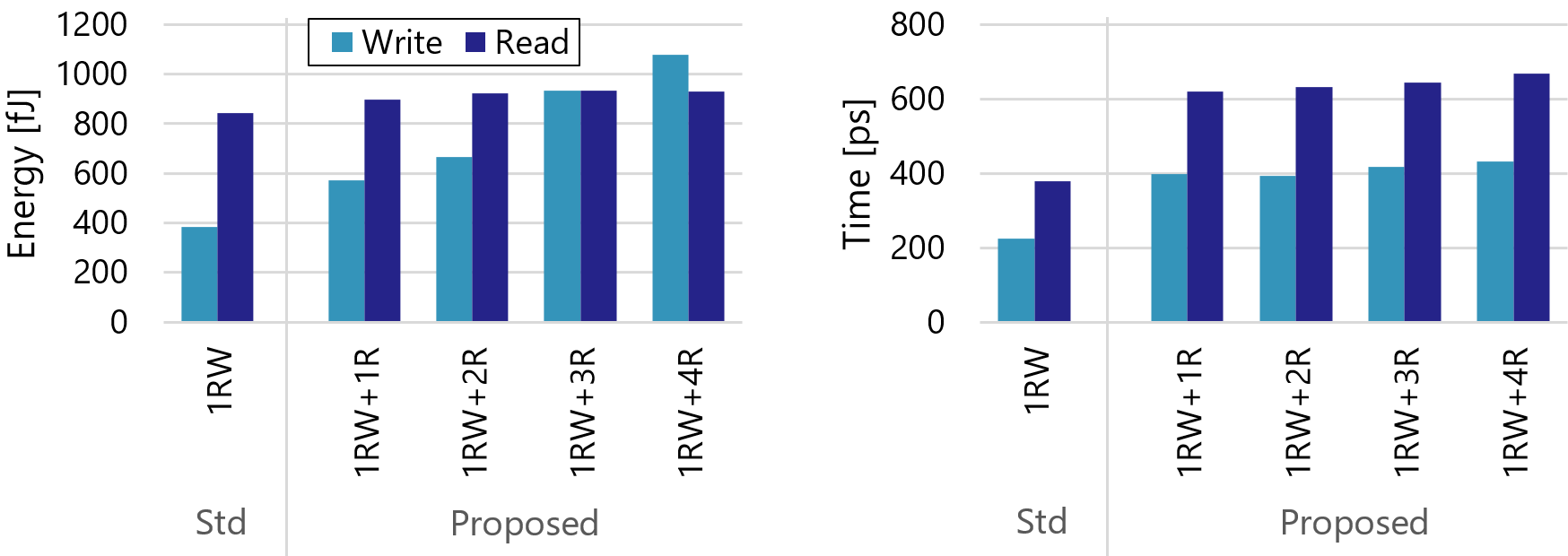}
    \vspace{-6mm}
    \caption{Write and Read Energies and Timings via Transposed port for different types of SRAM Cells.}
    \label{fig:writeRead}

    \vspace{10px}

    \includegraphics[width=\columnwidth]{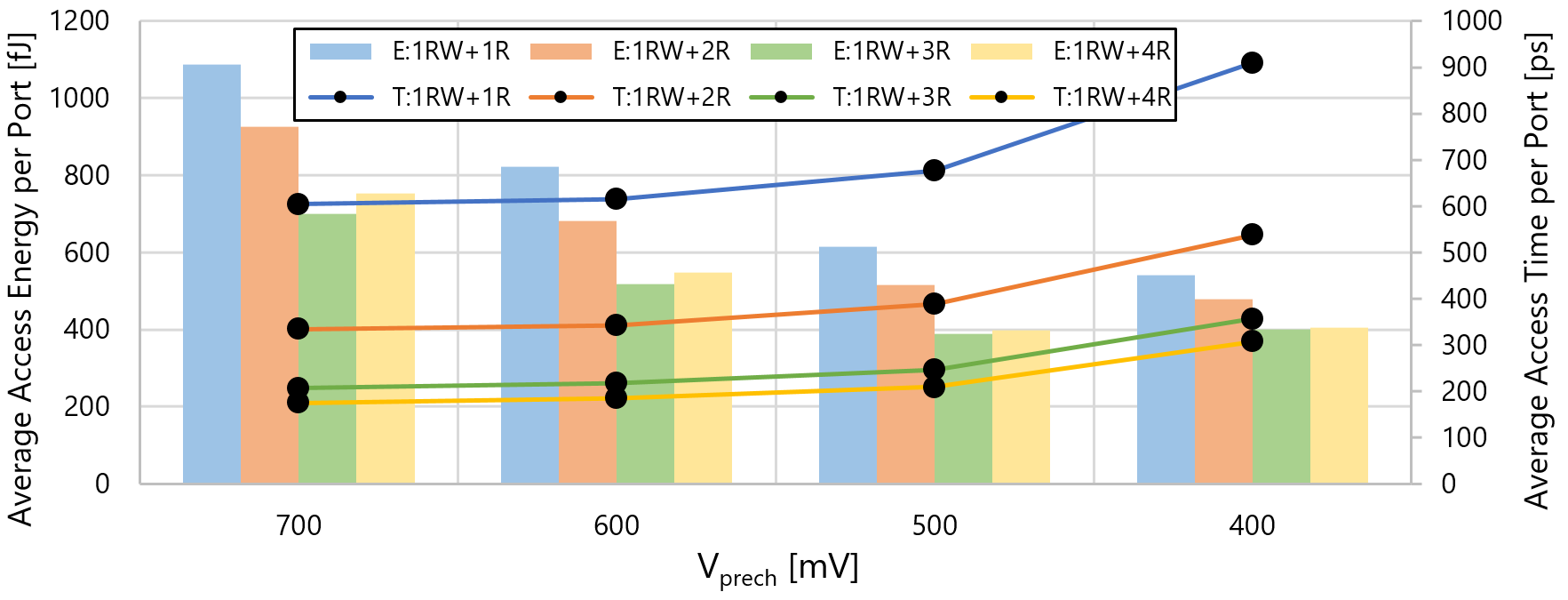}
    \vspace{-7mm}
    \caption{Illustration of the average access energy and time per number of ports for different $V_{prech}$ values and various ports for $128\times128$ arrays.}
    \label{fig:Vprech}
\end{figure}

Figure \ref{fig:writeRead} shows the time and energy measurements for Writing to the cell and Reading from the cell using the Transposed port (WL and BL/BLB). Write time is the delay between the start of the Write process and the cell content flipping to $90\%$ of its intended value. Read time is the delay between the Wordline being driven and the data output of the Sense Amplifier flipping. Additionally, Write energy is the energy consumed during the Write time, while Read energy is the energy consumed during a full clock cycle, which includes precharging of the BL/BLB \cite{SamanthaSRAM-TED20231}.

As expected, both the Write and Read operation results scale with the addition of ports due to the parasitics that these ports introduce. The effect is stronger for the Write operation, as the parasitics also require a lower value of $V_{WD}$ when more ports are added, increasing the voltage differential and consequently the power consumption. Note also that when just one extra Inference Port is added, there is an immediate and significant increase in both Write and Read times of the Transposed port. This is because the WL wire in the proposed cells is narrower and thus more resistive, which is necessary due to the new RBL0-RBL3 that have to be routed in the same metal layer.

Figure \ref{fig:Vprech} demonstrates the relationship between access time and energy consumption for different $V_{prech}$ levels and different numbers of ports in the SRAM. The results are shown for full utilization of all available ports, meaning that if a cell has $p$ ports, $p$ Read operations are performed. Total access time is calculated as the sum of the precharge time and the Read time.
\begin{itemize}
    \item \textbf{The effect of precharge voltage:} As Figure \ref{fig:Vprech} indicates, there is a trade-off between access time and energy. We select $V_{prech} = 500 mV$, as it leads to a reduction of at least $43\%$ in energy consumption at the cost of at most $19 \%$ higher access time for all port numbers. Lowering $V_{prech}$ from $500 mV$ to $400 mV$ saves up to $10\%$ more energy for 1- and 2-port designs. However, for 3- and 4-port designs energy consumption actually increases due to much slower precharging. 

    \item \textbf{The effect of the number of Inference ports:} Adding extra Inference ports increases the parallelism and reduces the average access time. However, the average access energy starts increasing after adding the fourth port. This increase in energy consumption per port can be explained by the increased cell size that creates more parasitics. These results support our earlier claim that increasing the ports to more than four would not be beneficial due to the excessive area and resulting parasitics.
\end{itemize}
\vspace{-2mm}
\subsection{Timing Evaluation}
To calculate the system clock period, we need to determine the duration of each pipeline stage.
Table \ref{tab:times} shows the durations of each stage for the various cell designs, including slack. The longest of the two stages determines the minimum clock period for each design. The Table shows that the critical path of the Arbiter does not scale with added ports. As a result, with more added ports the SRAM Read + Neuron accumulation stage becomes the bottleneck.

\begin{table}[b!]
\vspace{-4mm}
\caption{Time required for each stage in the pipeline for different SRAM cells, with the longest of the two stages indicating the clock period for each cell (1RW is the standard 6T cell).}
\label{tab:times}
\resizebox{0.9\columnwidth}{!}{
\begin{tabular}{l c c c c c}
\hline
 &
  \textbf{1RW} &
  \textbf{1RW+1R} &
  \textbf{1RW+2R} &
  \textbf{1RW+3R} &
  \textbf{1RW+4R} \\ \hline
\textbf{Arbiter} &
  1.01ns &
  1.01ns &
  1.04ns &
  1.03ns &
  1.01ns \\ \hline
\textbf{\begin{tabular}[c]{@{}l@{}}SRAM \\ + Neuron\end{tabular}} &
  0.69ns &
  1.08ns &
  1.18ns &
  1.14ns &
  1.23ns \\ \hline
\end{tabular}}
\end{table}
\vspace{-2mm}
\subsection{System-Level Evaluations} 

\subsubsection{Online Learning}
Online learning becomes significantly more efficient with Transposable memory access. 
Without the presence of the Transposed ports, reading and writing all the weights in an array of $128 \times 128$ 6T (1RW) SRAM cells would require $2 \times 128$ clock cycles, consuming 257.8 ns and 157 pJ. 
For our proposed SRAM cells, the synaptic weights of a post-synaptic Neuron can be read and written in just $2\times4$ cycles, where the factor 4 is due to the use of 4-to-1 MUXs. 
The 4-port cell, identified as the worst performer in the Transposed Read/Write port configuration, operates with a clock period of 1.2 ns.
This means it requires only $9.9$ ns ($26.0\times$ less) and $8.04$ ns ($19.5\times$ less) to read and write a full column.

\subsubsection{Inference}
To evaluate the system's Inference performance, we have created a Fully Connected Binary-SNN network for MNIST digit classification by placing multiple Tiles in sequence. The network has a structure of \textit{768:256:256:256:10}. In case a layer exceeds the maximum SRAM array size, multiple SRAM arrays are used in a single Tile. Each SRAM has its own 128-wide Arbiter. This increases parallelism by another factor; a 256-wide layer will have two $p$-port Arbiters, meaning up to $2\times p$ spikes can be selected per clock cycle. In order to reduce the input images from 784 to 768 pixels, a $2\times2$ set of pixels is removed from every corner of the images. This ensures that the first layer corresponds to exactly $6 \times 128$ inputs. We have trained the network as a \textit{Binary Neural Network} (BNN) with a sign activation function and per-neuron biases. The BNN is then converted to a Binary-SNN with per-neuron thresholds, as described in \cite{B-SNN-01-conversion}. The resulting accuracy achieved by the network is 97.64\%.

\begin{figure}[t!]
    \centering
    \includegraphics[width=0.75\columnwidth]{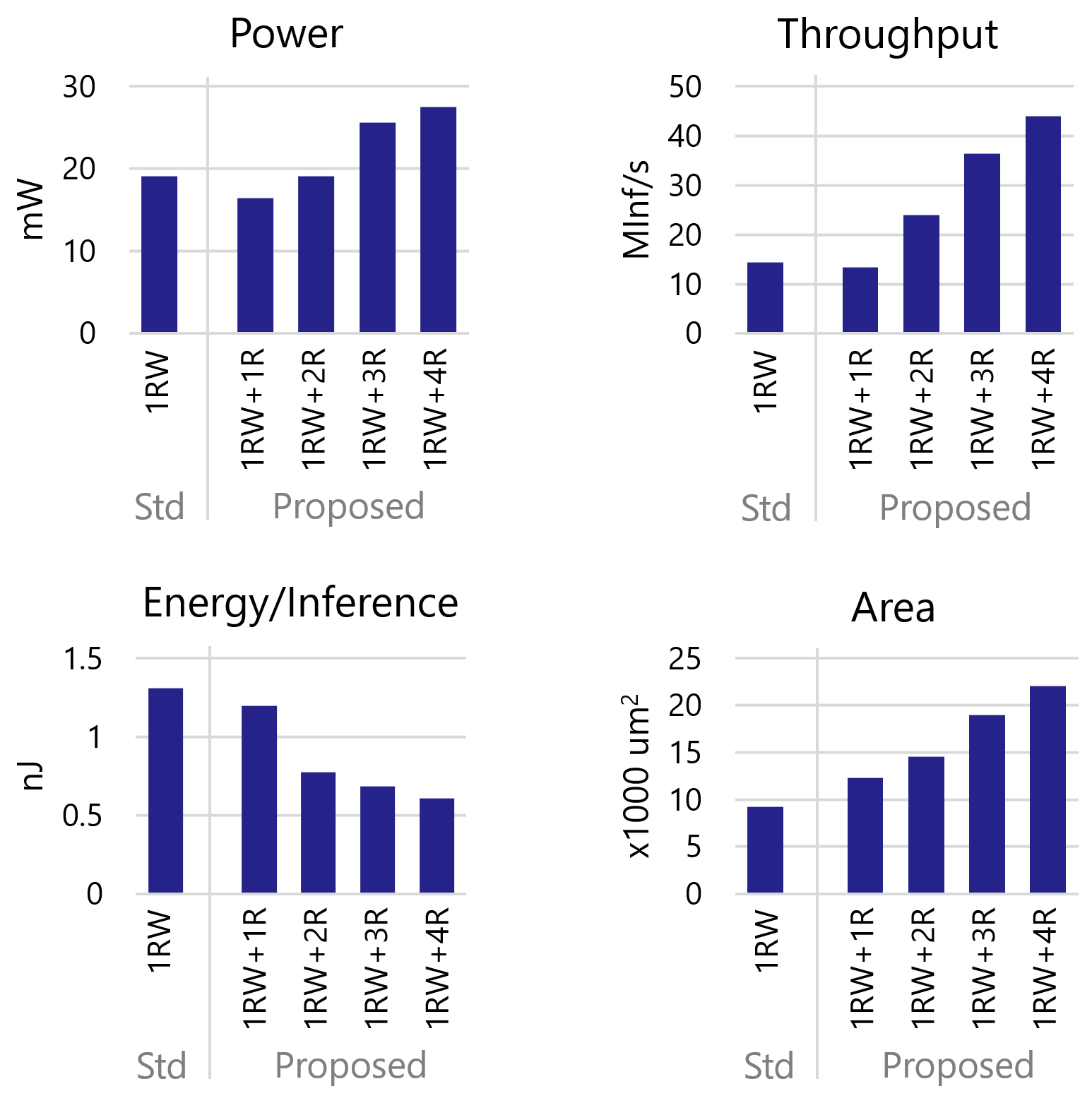}
    \vspace{-2mm}
    \caption{System-level comparison between using standard 6T (1RW) SRAM Cell and our proposed cells.}
    \label{fig:systemPortComparison}
    \vspace{-6mm}
\end{figure}

Figure \ref{fig:systemPortComparison} provides a comparison of system-level power, performance, energy, and area for the five discussed SRAM cell options. Generally, the power of the system increases with the number of added ports. However, the system's power implemented with the standard 1RW cells is higher than that of the 1RW+1R and 1RW+2R cells. This is due to the active power savings from the voltage scaling of $V_{prech}$ for the decoupled ports. When comparing the 1RW and 1RW+1R cells, throughput decreases slightly, as the effective parallelism is the same, but read operations for the 1RW+1R cell are slower due to the added parasitics. However, at 2+ added ports, the increased parallelism compensates for this. Additionally, with every added port, the overall energy/Inference decreases significantly due to the increased spike throughput. The main downside of the multiport design is that the area of the design using the 1RW+4R cell is $2.4\times$ larger than when using the standard 1RW cell.

Table \ref{tab:compTable} compares the 1RW+4R cell with state-of-the-art SNN accelerators aimed at ultra-low-power applications. We achieved typical overall power consumption, but significantly improve energy/Inference and throughput.
Note that the presented implementation is biased heavily towards high throughput. For applications that have lower throughput demands, a lower $VDD$, lower clock frequency, and HVT transistors can be utilized to significantly reduce power consumption, while maintaining similar energy/Inference.

\section{Conclusion}\label{sec:conclusion}
In our research paper, we presented an SRAM-based CIM accelerator designed for SNNs in IMEC's 3nm FinFET. The synthesis results indicate that our system exhibits high classification accuracy, while also reducing system-level power consumption and increasing the number of Inferences per second by four orders of magnitude. These accomplishments were made possible through the use of a new multiport SRAM cell that enables enhanced parallelism and voltage scaling, an improved spike arbitration system, a low-cost communication fabric, and technology scaling which is made possible by the fully digital design. 

\begin{table}[t!]
\caption{Comparison of the 1RW+4R-based system with state-of-the-art Small-scale SNN Accelerators.}
\label{tab:compTable}
\resizebox{\columnwidth}{!}{
\begin{tabular}{lllll}
\hline
\multicolumn{1}{|l|}{} &
  \multicolumn{1}{l|}{\textbf{\cite{mac-digital-1row}}} &
  \multicolumn{1}{l|}{\textbf{\cite{mac-digital-Nrow}}} &
  \multicolumn{1}{l|}{\textbf{\cite{transposable-barrelShift}}} &
  \multicolumn{1}{l|}{\textit{\textbf{This Work}}} \\ \hline
\multicolumn{1}{|l|}{\textbf{Technology {[}nm{]}}} &
  \multicolumn{1}{l|}{65} &
  \multicolumn{1}{l|}{10} &
  \multicolumn{1}{l|}{65} &
  \multicolumn{1}{l|}{3} \\ \hline
\multicolumn{1}{|l|}{\textbf{Neuron Count}} &
  \multicolumn{1}{l|}{650} &
  \multicolumn{1}{l|}{4096} &
  \multicolumn{1}{l|}{1K} &
  \multicolumn{1}{l|}{778} \\ \hline
\multicolumn{1}{|l|}{\textbf{Synapse Count}} &
  \multicolumn{1}{l|}{67K} &
  \multicolumn{1}{l|}{1M} &
  \multicolumn{1}{l|}{256K} &
  \multicolumn{1}{l|}{330K} \\ \hline
\multicolumn{1}{|l|}{\textbf{Activation Bit Width}} &
  \multicolumn{1}{l|}{6} &
  \multicolumn{1}{l|}{1} &
  \multicolumn{1}{l|}{--} &
  \multicolumn{1}{l|}{1} \\ \hline
\multicolumn{1}{|l|}{\textbf{Weight Bit Width}} &
  \multicolumn{1}{l|}{1} &
  \multicolumn{1}{l|}{7} &
  \multicolumn{1}{l|}{5} &
  \multicolumn{1}{l|}{1} \\ \hline
\multicolumn{1}{|l|}{\textbf{Transposable}} &
  \multicolumn{1}{l|}{No} &
  \multicolumn{1}{l|}{No} &
  \multicolumn{1}{l|}{Yes} &
  \multicolumn{1}{l|}{Yes} \\ \hline
\multicolumn{1}{|l|}{\textbf{Clock Frequency}} &
  \multicolumn{1}{l|}{70kHz} &
  \multicolumn{1}{l|}{506MHz} &
  \multicolumn{1}{l|}{100MHz} &
  \multicolumn{1}{l|}{810MHz} \\ \hline
\multicolumn{5}{|l|}{{\textit{\textbf{MNIST}}}} \\ \hline
\multicolumn{1}{|l|}{\textbf{Power}} &
  \multicolumn{1}{l|}{305nW} &
  \multicolumn{1}{l|}{196mW*} &
  \multicolumn{1}{l|}{53mW} &
  \multicolumn{1}{l|}{29.0mW} \\ \hline
\multicolumn{1}{|l|}{\textbf{Accuracy {[}\%{]}}} &
  \multicolumn{1}{l|}{97.6} &
  \multicolumn{1}{l|}{97.9} &
  \multicolumn{1}{l|}{97.2} &
  \multicolumn{1}{l|}{97.6} \\ \hline
\multicolumn{1}{|l|}{\textbf{Throughput {[}inf/s{]}}} &
  \multicolumn{1}{l|}{2} &
  \multicolumn{1}{l|}{6250} &
  \multicolumn{1}{l|}{20} &
  \multicolumn{1}{l|}{44M} \\ \hline
\multicolumn{1}{|l|}{\textbf{Energy/Inf {[}nJ{]}}} &
  \multicolumn{1}{l|}{195} &
  \multicolumn{1}{l|}{1000} &
  \multicolumn{1}{l|}{--} &
  \multicolumn{1}{l|}{0.607} \\ \hline
\multicolumn{5}{l}{\begin{tabular}[c]{@{}l@{}}* Inferred from SOP/s/mm\textsuperscript{2}, Area, and pJ/SOP
\end{tabular}}
\end{tabular}}
\vspace{-4mm}
\end{table}



\bibliographystyle{ACM-Reference-Format}

\end{document}